\documentclass[10pt,english,two column]{IEEEtran}
\usepackage[T1]{fontenc}
\usepackage[latin9]{inputenc}
\usepackage{array}
\usepackage{amsmath}
\usepackage{amssymb}
\usepackage{graphicx}

\makeatletter

\providecommand{\tabularnewline}{\\}

\ifCLASSINFOpdf
\else
\fi
\usepackage{babel}

\usepackage{eso-pic}

\AddToShipoutPicture{%
  \AtPageUpperLeft{%
    \hspace*{20pt}\makebox(200,-20)[lt]{%
      \footnotesize%
      \textbf{TO APPEAR IN IEEE TRANSACTIONS ON INTELLI. TRANSP. SYST., 2016}%
}}}

\makeatother

\usepackage{babel}
\begin{document}

\title{Topology Discovery for Linear Wireless Networks with Application
to Train Backbone Inauguration}

\author{Yu Liu, Jianghua Feng, Osvaldo Simeone,\IEEEmembership{\ Senior Member, IEEE,}
Jun Tang, Zheng Wen, Alexander M. Haimovich,~\IEEEmembership{Fellow, IEEE,}
MengChu Zhou,~\IEEEmembership{Fellow, IEEE} \thanks{Y. Liu, O. Simeone, A. M. Haimovich and M. Zhou are with the Center
for Wireless Communications and Signal Processing Research (CWCSPR),
ECE Department, New Jersey Institute of Technology (NJIT), Newark,
USA (email: \{yl227, osvaldo.simeone, haimovic, zhou\}@njit.edu). 

J. Feng, J. Tang and Z. Wen are with CSR Zhuzhou Institute Co., LtD,
Shidai Road, Zhuzhou, Hunan Province, China. E-mail: \{fengjh, tangjun,
wenzheng\}@csrzic.com.}}
\maketitle
\begin{abstract}
A train backbone network consists of a sequence of nodes arranged
in a linear topology. A key step that enables communication in such
a network is that of topology discovery, or train inauguration, whereby
nodes learn in a distributed fashion the physical topology of the
backbone network. While the current standard for train inauguration
assumes wired links between adjacent backbone nodes, this work investigates
the more challenging scenario in which the nodes communicate wirelessly.
The key motivations for this desired switch from wired topology discovery
to wireless one are the flexibility and capability for expansion and
upgrading of a wireless backbone. The implementation of topology discovery
over wireless channels is made difficult by the broadcast nature of
the wireless medium, and by fading and interference. A novel topology
discovery protocol is proposed that overcomes these issues and requires
relatively minor changes to the wired standard. The protocol is shown
via analysis and numerical results to be robust to the impairments
caused by the wireless channel including interference from other trains. 
\end{abstract}

\section{Introduction\label{sec:Introduction}}

\begin{figure}[htbp]
\begin{centering}
\textsf{\includegraphics[width=8.8cm,height=3.9cm]{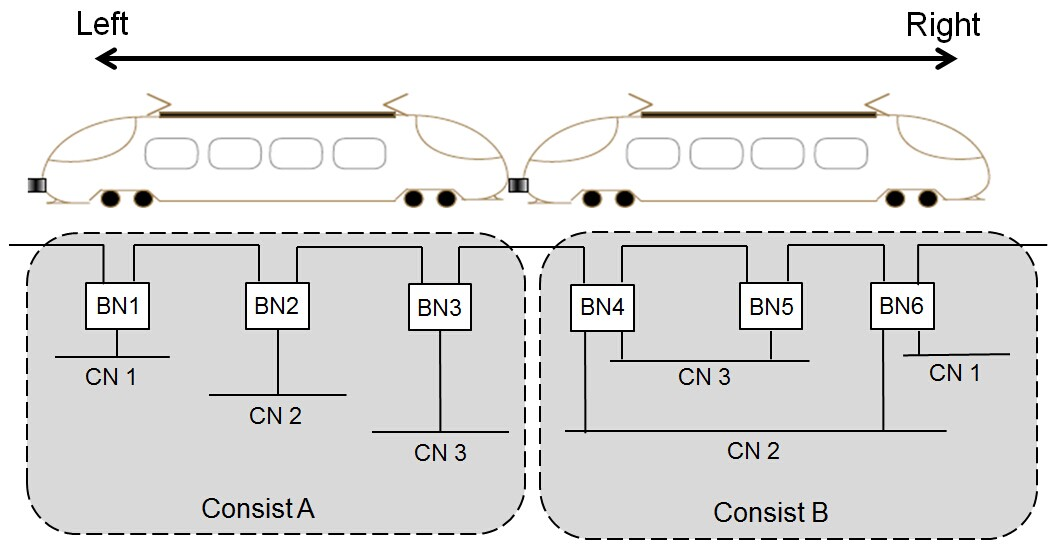}}
\par\end{centering}

\caption{\label{fig. wiredTrain} A wired backbone network with indicated physical
and logical topologies. The physical topology refers to the ordering
of the backbone nodes (BNs) and the logical topology means the ordering
of the consist networks (CNs).}
\end{figure}

Wireless networks in which radio nodes are deployed according to a
linear topology find numerous applications, including inter-vehicle
communication systems \cite{Kesting}-\cite{CHen} and train backbone
communications \cite{Ning}, \cite{IEC 61375} (see Fig. \ref{fig. wiredTrain}).
For such networks, it is convenient to have an automatic procedure
that learns the network topology for any given configuration both
at power-up time and in case nodes are added, replaced or removed
during the network operation. For instance, the neighboring car in
a train backbone is subject to change due to the fact that a single
car or a group of cars of a train may be separated during the operation
of shortening or lengthening a train. As a result, it is desirable
that the system can learn the network topology when powered up and
update any change in the topology as they occur without the intervention
of a human operator. This is done via the process of topology discovery
(TD).

The basic task of TD is to enable each node to learn the physical
topology of the network. The physical topology consists of an ordered
list of the media access control (MAC) addresses of the nodes in the
network, where the order reflects the physical location of the nodes
in the linear topology. According to current standards \cite{IEC 61375},
the process operates via the exchange of MAC-level messages among
the nodes in a distributive fashion. To illustrate the concept of
a physical topology, an example is provided in Fig. \ref{fig. wiredTrain}
for a train backbone network. In this example, the physical topology
lists the MAC addresses of the nodes, referred to as backbone nodes
(BNs), in the order from 1 to 6\footnote{The starting point of the ordering of the nodes in the physical topology
is fixed at the time of deployment.}. 

The TD protocol (TDP) \cite{IEC 61375} that is currently being standardized
for train backbone communications applies to wired train backbone
networks, in which the BNs are connected to their neighbors via dedicated
wires. The TDP consists of two phases: 1) \emph{neighbor discovery}:
in this phase, each BN finds the MAC address of the neighboring BNs;
2) \emph{topology discovery}: in this phase, the physical topology
is detected\footnote{The standard also considers the discovery of the ``logical'' topology
of the train, which will be discussed in Sec. \ref{sec:Wired Train TD}.} via message exchange at the MAC layer. To implement TDP, the BNs
transmit two types of MAC frames \cite{IEC 61375}: 1) \textit{hello
frames}, which carry only the MAC address of the sender BN and are
used for neighbor discovery; and 2) \textit{topology frames}, which
carry information about the MAC addresses of the BNs currently ``discovered''
by the sender BN and are used for topology discovery. 

While the standard \cite{IEC 61375} applies to wired backbone networks,
there is high interest in the industry to develop a fully wireless
solution. The key motivations for this desired switch from wired topology
discovery to wireless one are the flexibility and the capability for
expansion and upgrading of a wireless backbone. However, as it will
be discussed, the wired TDP \cite{IEC 61375} does not lend itself
to an implementation with wireless nodes. Moreover, any wireless implementation
needs to contend with the inherently less reliable transmission medium.
This work is hence devoted to developing a novel TDP, which will be
referred to as wireless TDP (WTDP), that builds on the standard \cite{IEC 61375}
but is suitable for implementation on a wireless backbone. Specifically,
the aim of the proposed WTDP is to retain the main features of the
wired counterpart TDP \cite{IEC 61375}, while adapting messages and
protocols to the new requirements for a wireless implementation. 

\begin{figure}[htbp]
\begin{centering}
\textsf{\includegraphics[width=8.8cm,height=2cm]{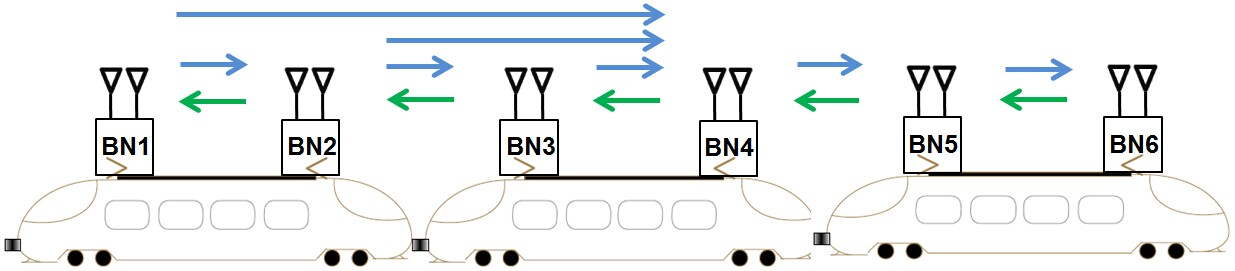}}
\par\end{centering}

\caption{\label{fig. Wireless1}A wireless backbone network with indication
of signals interfering the reception of BN 4.}
\end{figure}

The implementation of TDP \cite{IEC 61375} over a wireless network
is made difficult by the broadcast nature of the wireless medium,
and by fading and interference. Consider for instance the neighbor
discovery phase. In wired TDP, hello frames are transmitted only to
the neighbor(s) of a node as shown in Fig. \ref{fig. wiredTrain}.
The neighbor discovery phase hence only requires that a single hello
frame be received correctly from each neighbor. Wireless broadcasting
instead, causes a frame to be received also by BNs that are not physical
neighbors, making the detection of physical neighbors challenging
\cite{Zeng}. This effect is compounded by the fact that, due to fading
and interference, there is a non-zero probability that decoding errors
impair the transmission from physical neighbors more significantly
than the transmission from further nodes. Unlike the wired case, simultaneous
transmissions in the same frequency band may lead to interference,
which may cause the loss of a packet. For instance, with reference
to Fig. \ref{fig. Wireless1}, it is possible for BN 4 to decode the
hello frame sent by BN 2 correctly, while decoding the hello frame
from BN 3 incorrectly due to fading or interference. Another issue
is that the standard \cite{IEC 61375} prescribes the multicasting
of a topology frame to all the BNs in the network. In a wireless implementation,
this is bound to create large backlogs and excessive interference.
The proposed WTDP, detailed later in Sec. \ref{sec:Proposed-Inauguration-Scheme},
aims to address these challenges. 

We conclude this section with a remark on related work. Studies on
wireless network topology discovery include \cite{Zeng}-\cite{MZhang}
and references therein. In them, the key underlying assumption is
that two nodes are considered to be neighbors if they are within their
respective transmission ranges such that it is possible to establish
a direct link between them. The topology discovery protocol hence
aims at identifying connectivity, or reachability, properties of the
network. This is typically done either by checking if a hello message
is successfully received \cite{Zeng}-\cite{LiuX} or by measuring
received signal strengths \cite{IJawhar}-\cite{MZhang}. The design
of specific topology discovery algorithms has been conducted in the
context of different protocols such as IEEE 802.11, e.g., \cite{Hermann},
or ZigBee \cite{HXie}. Note that unlike the works in \cite{B-ning-1}
and \cite{B-ning-2}, wherein the inter-train and train-ground communication
problems are addressed, this paper focuses on the problem of initializing
the network for intra-train communication. 

The topology discovery schemes that are available in the literature
do not solve the problem of interest for the train backbone. The key
distinguishing feature is that, in classical topology discovery, as
discussed, a node is considered to be a neighbor as long as it is
reached with a significantly large power. This goal is completely
different from the requirements of train backbone inauguration, in
which instead a neighbor is defined by its physical location and not
by the strength of the received power. To see the difference, note
that each backbone node has only two neighbors, one that should be
specified as left-neighbor and one as right-neighbor. In contrast,
a classical topology discovery scheme may identify an arbitrary number
of neighbors that happen to receive the transmitted signal with sufficient
power without consideration of their physical location. The approach
proposed in this paper is meant to address the errors that can arise
with conventional topology discovery schemes whereby physical neighbors
may be incorrectly detected due to the fact that they receive a signal
with sufficient strength.

The rest of the paper is organized as follows. In Sec. \ref{sec:Wired Train TD},
we review the TDP standard \cite{IEC 61375}, while in Sec. \ref{sec:Proposed-Inauguration-Scheme},
we detail the proposed WTDP. In Sec. \ref{sec:Performance-Analysis-with},
we provide a performance analysis for the neighbor discovery phase
of the proposed WTDP implemented with a slotted ALOHA MAC protocol.
In Sec. \ref{sec:ND-parellel_tracks}, we describe a case study consisting
of two parallel trains. Numerical results of the proposed WTDP along
with the performance analysis of neighbor discovery are presented
in Sec. \ref{sec:Numerical-Results-and}.

\section{Background: Wired Train Topology Discovery\label{sec:Wired Train TD}}

In this section, we briefly review the standard wired TDP \cite{IEC 61375}.
We observe that TD is also referred to as inauguration in \cite{IEC 61375}.
Before the inauguration process, each BN knows its own MAC address
and also the unique identifier (ID) of the consist networks (CNs)
that are connected to the BN. A CN represents a subnetwork on the
train. BNs may belong to multiple CNs, as illustrated in Fig. \ref{fig. wiredTrain}.
The goal of TDP is to enable all the BNs to learn the physical and
logical topologies of the train. As mentioned, the physical topology
consists of an ordered list of BNs. The logical topology refers to
an ordered list of CNs, with indication for each CN of the participant
BNs, where the order reflects the physical location of the CNs. For
instance, the logical topology for the network in Fig. \ref{fig. wiredTrain}
lists the CN IDs in the order A.1, A.2, A.3, B.3, B.2 and B.1, along
with the corresponding MAC address of the BNs, as shown in Table.
\ref{tab:phy=000026logi toplgy-table}. After inauguration, a BN ID
is assigned to each BN according to the identified physical topology,
and a subnet ID is assigned to each CN following the logical topology
that is discovered. Taking the backbone network in Fig. \ref{fig. wiredTrain}
as an example, all six BNs are assigned with BN IDs in the ascending
order from the left end to the right end, as illustrated in Table
\ref{tab:phy=000026logi toplgy-table}. In the same order, the subnet
IDs are assigned according to the logical topology (see \cite{IEC 61375}
for further details).

\begin{table}[tp]
\caption{\label{tab:phy=000026logi toplgy-table}Physical and logical topologies}

\centering{}%
\begin{tabular}{|c|>{\centering}p{2.5cm}|>{\centering}p{1.7cm}|>{\centering}p{1.7cm}|}
\hline 
CN ID & MAC address of connected BN & assigned BN ID & assigned subnet ID\tabularnewline
\hline 
\hline 
A.1 & BN 1's MAC address  & 1 & 1\tabularnewline
\hline 
A.2 & BN 2's MAC address  & 2 & 2\tabularnewline
\hline 
A.3 & BN 3's MAC address  & 3 & 3\tabularnewline
\hline 
B.3 & BN 4's MAC address  & 4 & 4\tabularnewline
\hline 
B.3 & BN 5's MAC address  & 5 & 4\tabularnewline
\hline 
B.2 & BN 4's MAC address  & 4 & 5\tabularnewline
\hline 
B.2 & BN 6's MAC address  & 6 & 5\tabularnewline
\hline 
B.1 & BN 6's MAC address  & 6 & 6\tabularnewline
\hline 
\end{tabular}
\end{table}

Each BN, except the two at the beginning and end of the train, has
two outgoing links, one toward its neighbor to the ``right'' and
one towards the ``left''. Note that the notions of ``left'' and
``right'' are common to all BNs on the backbone and are set by construction.
Similarly, each BN has also two incoming links, one from the neighbor
on the left and one from the neighbor on the right. As can be seen
in Fig. \ref{fig. wiredTrain}, there are then an outgoing and an
incoming link between a BN and a neighbor.

The BNs send two different types of frames, namely hello frames and
topology frames. Both hello and topology frames are transmitted periodically
and continuously. The hello frame contains the MAC address of the
sender BN. This frame is sent only to the nodes' neighbors. The topology
frame instead contains information about the MAC addresses of previously
discovered nodes. Specifically, the topology frame sent to the neighbor
on the right contains an unordered list of all the currently known
MAC address of the BNs on the left of the BN, and vice versa for the
topology frame sent to the neighbor on the left. Topology frames are
to be forwarded by the receiving BN in the same direction they have
been received. This way, a topology frame is multicast to all BN in
the given direction. As an example, in Fig. \ref{fig. wiredTrain},
if BN 4 has recognized BN 3 as a neighbor and has discovered that
BNs 5 and 6 are on its right, the topology frame sent by BN 4 to BN
3 includes an unordered list of the MAC addresses of BNs 5 and 6.
The topology frame also contains the IDs of the CNs that are connected
to the sender BN, namely, CN B.3 and CN B.2 are connected to BN 4.
After BN 3 receives this topology frame, the frame is forwarded to
BN 2.

To summarize, the operation of TDP can be divided into two conceptually
different phases, namely neighbor discovery and topology discovery.
\begin{itemize}
\item \emph{Neighbor discovery}: Each BN receives hello frames from its
incoming links to the left and/or to the right. Since each of these
frames carries the MAC address of the sending neighbor, the BN at
hand learns the MAC address of its neighbors after receiving one frame
from the left and one from the right. The reception of these two frames
completes the neighbor discovery phase.
\item \emph{Topology discovery}: Each BN keeps updated physical and logical
topology tables (see Table \ref{tab:phy=000026logi toplgy-table})
based on the previously received hello and topology frames. As mentioned
above, each transmitted topology frame to the left/right contains
an unordered list of MAC addresses and the CN IDs that are connected
to the sender BN. Upon reception of a topology frame, a BN updates
its physical and logical topology tables. The BN can also check on
whether its current tables coincide with the ones available at the
BN that produced the topology frame thanks to a cyclic redundancy
check (CRC) included in each topology frame.
\end{itemize}
Inauguration is declared to be complete by an operator that has access
to the outcomes of the CRC steps carried out by the BNs.

\section{WTDP: Wireless Topology Discovery Protocol\label{sec:Proposed-Inauguration-Scheme}}

As discussed in Sec. \ref{sec:Introduction}, the wireless implementation
of TDP poses significant technical challenges. To overcome these problems,
the proposed WTDP prescribes a number of design choices at the deployment
and protocol level as discussed in this section.

\subsection{Deployment\label{sub:Deployment}}

WTDP is based on a physical implementation of the system that leverages
directional antennas and frequency planning.
\begin{itemize}
\item \textit{Directional antenna}: All the BNs have two directional antennas
and share the notion of a ``left'' and a ``right'' direction.
Each BN hence can transmit and receive on both its right-pointing
and left-pointing antennas. Note that the assumption concerning the
common notion of the left and right directions is consistent with
the model considered in the wired standard \cite{IEC 61375}. Directional
antennas enable a BN to distinguish between the signals received from
the left and right directions. 
\item \textit{Frequency planning}: To cope with interference, two sets of
frequencies are used, one for the right-pointing antennas and one
for the left-pointing antennas. Each directional antenna operates
on two different frequencies, one for transmission and one for reception.
Moreover, the same frequency is reused every $F$ hops. Therefore,
if $F=1$, we have full frequency reuse in each direction; instead,
if $F>1$, there are $F-1$ BNs transmitting in the same direction
but using different frequencies between two transmitters using the
same frequency. We refer to Fig. \ref{fig:Deployment} for an illustration.
Note that, with a frequency reuse $1/F$, the closest non-neighboring
BN that may receive a hello frame is $F-1$ hops away. A more conservative
frequency reuse hence reduces the danger of receiving a hello frame
from a non-neighboring BN. A smaller frequency reuse also reduces
the effect of interference.
\begin{figure}[htbp]
\begin{centering}
\textsf{\includegraphics[width=8.8cm,height=1.8cm]{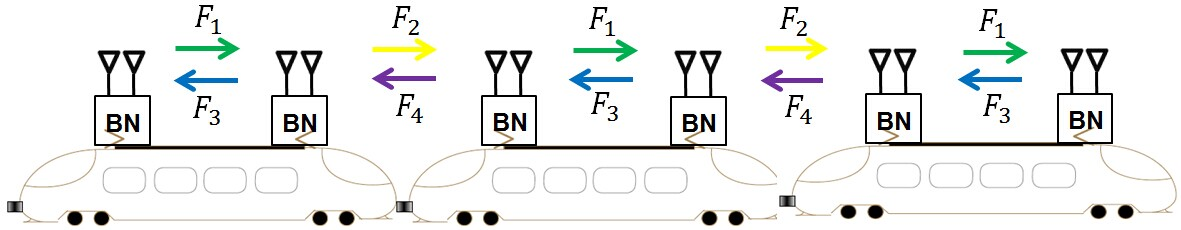}}
\par\end{centering}

\caption{\label{fig:Deployment}Illustration of the proposed physical implementation
of WTDP with directional antenna and frequency planning with $F=2$.
$F_{i}$ indicates the $i$th available carrier frequency.}
\end{figure}

\end{itemize}

\subsection{The Proposed Protocol}

\begin{figure*}[tbph]
\begin{centering}
\textsf{\includegraphics[width=15cm,height=2.5cm]{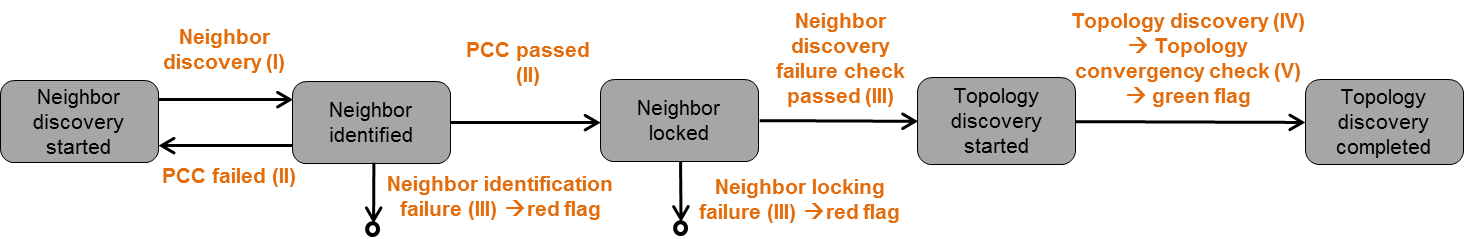}}
\par\end{centering}

\caption{\label{fig:highLevelFlow}A high-level illustration of the operation
of a BN in the proposed WTDP. The circle indicates that a red flag
has been raised and that the BN is waiting for the inauguration process
to be restarted. }
\end{figure*}

\begin{figure*}[tp]
\begin{centering}
\textsf{\includegraphics[width=17.5cm,height=12cm]{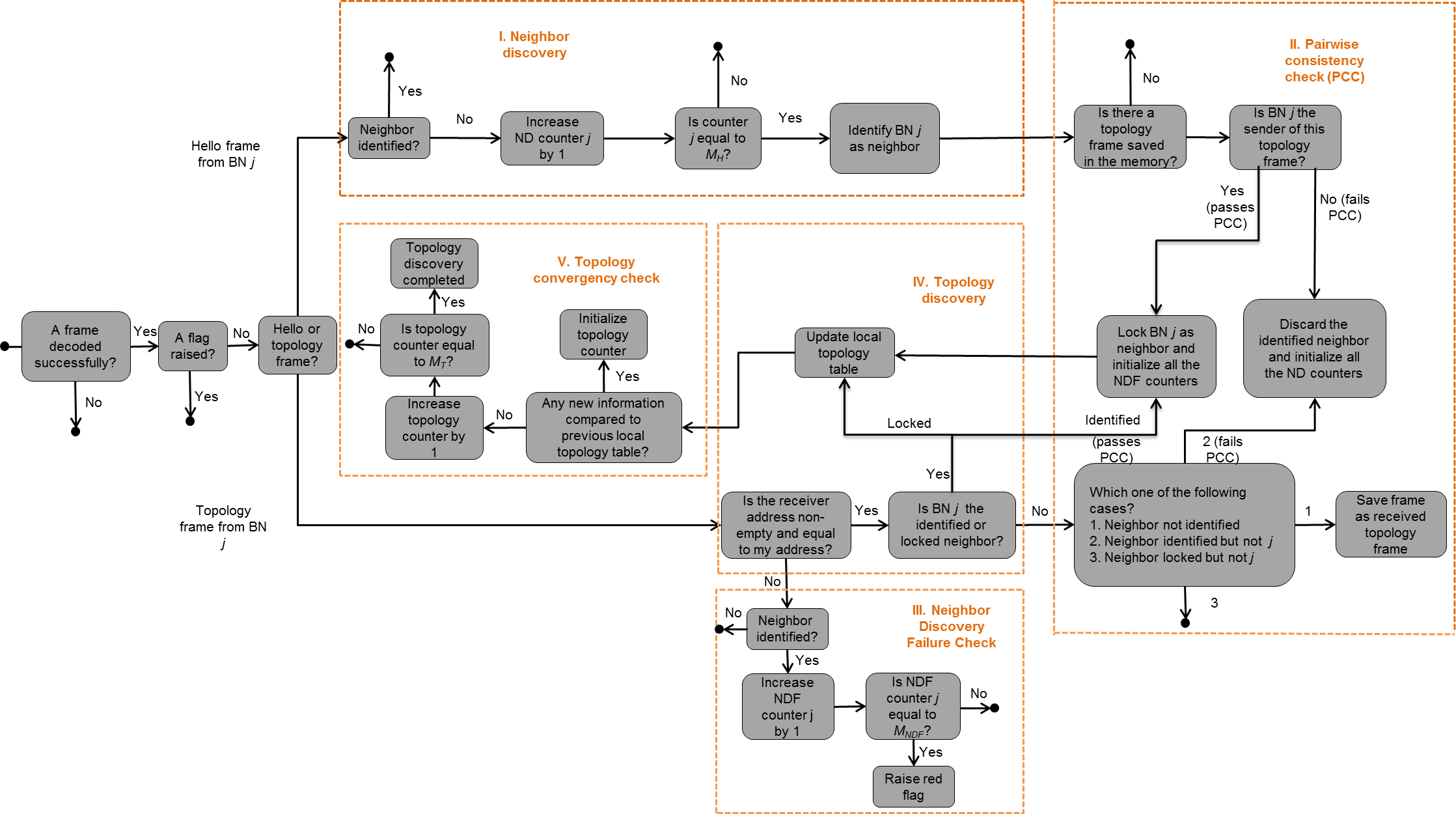}}
\par\end{centering}

\caption{\label{fig:flow}Illustration of the operation of a BN in the proposed
WTDP. The black dot means that the BN is ready to process the next
frame.}
\end{figure*}

In this section, we detail the operation of the proposed WTDP. The
proposed WTDP prescribes each BN to operate according to the high-level
flowchart of Fig. \ref{fig:highLevelFlow}, which is further detailed
in Fig. \ref{fig:flow}. As shown in Figs. \ref{fig:highLevelFlow}
and \ref{fig:flow}, the proposed WTDP includes five phases: I. Neighbor
discovery; II. Pairwise consistency check (PCC); III. Neighbor discovery
failure check; IV. Topology discovery; and V. Topology convergency
check. We first explain the flow of these phases with reference to
Fig. \ref{fig:highLevelFlow}. We then detail the steps of the algorithm
in the following subsections.

\subsubsection{Overview\label{sub:over}}

At first, each BN attempts to identify a neighbor during the neighbor
discovery phase (I). If a neighbor is identified, then, in order to
correct some of the possible errors of neighbor discovery errors,
the BN performs the PCC (II). If the identified neighbor passes the
PCC, it is upgraded to a locked neighbor; otherwise the identified
neighbor is discarded and neighbor discovery needs to be restarted.
After an identified neighbor is established, the BN also starts the
neighbor discovery failure check phase (III). Whenever a neighbor
discovery failure is detected, a ``red flag'' is raised. Upon the
observation of a red flag, the operator, which is informed about ``green
flags'' or ``red flags'' raised by the BNs, should restart the
inauguration process. Once a locked neighbor is established, the topology
discovery phase (IV) starts. The completion of the topology discovery
phase for each BN is indicated by the BN via a raised ``green flag'',
which signals that the topology convergency check (V) is passed.

\subsubsection{Data Structures\label{sub:Data-Structures}}

Each BN stores and updates the following data structures, whose use
will be detailed in the next subsections. 
\begin{itemize}
\item \textit{Neighbor discovery (ND) counters}: a list of neighbor discovery
counters, one for each of the BNs from which the BN has received a
hello frame;
\item \textit{Topology table}: an ordered list of the MAC addresses of the
discovered BNs, where the order reflects the physical location of
the BNs;
\item \textit{Topology counter}: a counter that accounts for the current
number of consecutive times that a topology frame has been received
but the local topology table has not been changed;
\item \textit{Neighbor discovery failure (NDF) check counters}: a list of
counters, one for each of the BNs from which a topology frame addressed
to any other BN has been received.
\end{itemize}

\subsubsection{Neighbor Discovery\label{sub:Neighbor-Discovery}}

As discussed in Sec. \ref{sec:Introduction}, neighbor discovery in
a wireless train backbone is significantly more complex than in the
wired counterpart system. This is due to the broadcast properties
of the wireless channel, which cause the hello frame transmitted by
a BN to be received not only by the actual neighbor BN but generally
also by further away BNs. As a result, unlike in the wired system,
reception of the hello frame does not, per se, establish that the
sender BN is a neighbor. 

In order to achieve neighbor discovery, the proposed scheme leverages
the fact that, on the\emph{ average, }the power received from an actual
neighboring BN is larger than that received from any other BN. This
is due to the lower path loss between closer nodes. Therefore, for
instance, it is more likely that a hello frame is received correctly
from an actual neighbor than from farther BNs. It is critical to note,
however, that, due to fading, it cannot be excluded that a hello frame
from a non-neighboring BN is received successfully, while that of
the actual neighbor is not. 

Based on the discussion above, we propose the following simple neighbor
discovery algorithm. For each hello frame correctly decoded in either
direction, if the MAC of the sender BN is already in the list of ND
counters, then the corresponding counter is increased by one; else,
a new counter is created, initialized to zero and associated to the
MAC address at hand. A node is identified to be a neighbor if it is
the first whose ND counter reaches a pre-defined threshold $M_{H}$.
In this event, this node is defined as the \textit{identified neighbor}
of the receiving BN. The described operations are within in the ``neighbor
discovery'' block of Fig. \ref{fig:flow}.

We remark that the simple algorithm proposed above makes exclusive
use of information available at the MAC layer, namely the number of
successfully received frames from different MAC addresses. This choice
has been made in order to allow for a simpler implementation, and
is in line with the wired counterpart standard.

\subsubsection{Pairwise Consistency Check (PCC) \label{sub:Pairwise-Consistency-Check}}

In order to reduce the probability of incorrect neighbor discovery,
we propose to perform a pairwise consistency check (PCC) upon the
reception of a topology frame. The key observation is that the topology
frame is addressed to the currently identified neighbor. Note that
the hello frames are instead broadcast. Therefore, based on the reception
of topology frames, each BN can verify whether the neighbor discovery
is pairwise consistent with respect to its neighbor in either direction.
By pairwise consistency, we mean that two BNs consider each other
as neighbors, one on the left and the other on the right. If a topology
frame is received from a BN that is not considered as a neighbor,
then the receiving BN can conclude that neighbor discovery is not
pairwise consistent in the direction of the received packet. 

To be specific, if the topology frame is received from the currently
identified neighbor, this identified neighbor passes the PCC and is
upgraded to the status of \textit{locked neighbor}. Once a locked
neighbor is established for a BN, any received topology frame from
other BNs is discarded. Instead, if a BN receives a topology frame
from a BN different from the identified neighbor, its identified neighbor
fails the PCC and all ND counters are reinitialized to zero in order
to restart the neighbor discovery phase for the receiving BN. Note
that, if a topology frame is received before any identified neighbor
is established, the frame will be saved for a PCC later. The detailed
procedure for PCC is described within the ``pairwise consistency
check'' block of Fig. \ref{fig:flow}.

\subsubsection{Neighbor Discovery Failure Check \label{sub:NDFC}}

\begin{figure}[htbp]
\begin{centering}
\textsf{\includegraphics[width=7.8cm,height=1.6cm]{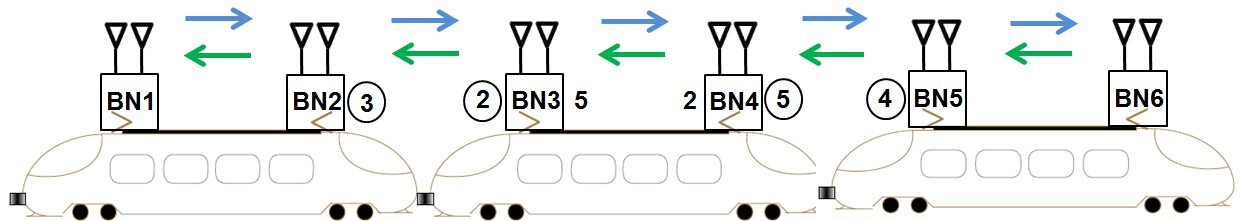}}
\par\end{centering}

\caption{\label{fig:PCC}An example of neighbor identification failure. A number
on the left/right of each BN refers to identified neighbors on the
left/right of the BN. A circled numbers means a neighbor has been
locked. }
\end{figure}

PCC helps improve the accuracy of neighbor discovery, but it does
not rectify errors that occur when two neighboring BNs identify their
neighbors incorrectly. This type of failure is defined as neighbor
identification failure. An example is shown in Fig. \ref{fig:PCC}.
It is seen that, if BN 3 identifies BN 5 as a neighbor and BN 4 identifies
BN 2 as a neighbor, this error cannot be corrected by PCC because
neither BN 3 nor BN 4 will send a topology frame to the other. 

In order to identify the neighbor discovery failure described above,
we propose to perform neighbor discovery failure check. The idea is
that, after a neighbor has been identified but not locked, if a BN
receives too many topology frames addressed to a BN other than itself,
it is probable that its actual neighbor had identified some other
BN as its neighbor. In this case, this BN cannot successfully complete
neighbor discovery and a red flag is raised. Specifically, each BN
maintains an NDF counter, which counts the number of topology frames
addressed to other BNs that are received after a neighbor has been
identified. When the NDF counter reaches a pre-defined threshold $M_{NDF}$,
the BN raises a red flag warning the train operator of a neighbor
discovery failure. 

\begin{figure}[htbp]
\begin{centering}
\textsf{\includegraphics[width=8.8cm,height=1.6cm]{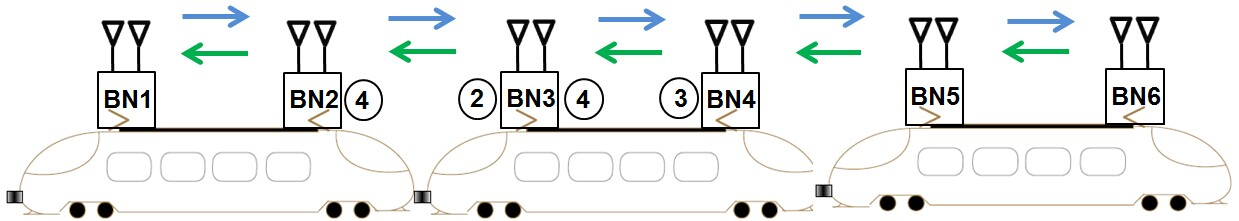}}
\par\end{centering}

\caption{\label{fig:PCC2}An example of neighbor locking failure. A circled
number on the left/right of each BN refers to a locked neighbor on
the left/right of the BN.}
\end{figure}

The other possible neighbor discovery failure happens when a BN is
established as the locked neighbors by more than one BN. This causes
the problem that certain BNs do not receive topology frame from their
locked neighbors and thus topology discovery will never be completed.
We define this type of failure as neighbor locking failure. An example
is shown in Fig. \ref{fig:PCC2}, where although BN 2 established
BN 4 as its locked neighbor, BN 4 has locked with BN 3, and hence
no topology frames will be received by BN 2 from BN 4. To deal with
this problem, after the neighbor of a BN is locked, the NDF counter
is initialized and used to count topology frames received from BNs
different from its locked neighbor. A red flag is raised if the NDF
counter exceeds the threshold. 

We remark that due to the introduction of the phase II (PCC) and phase
III (neighbor discovery failure check), the proposed protocol is a
bidirectional protocol. Hence, these phases can also be used to counteract
hello flooding attacks\footnote{In a hello flooding attack, hello messages/frames are transmitted
or tunneled with a very abnormal high power convincing many surrounding
nodes that the malicious node is their neighbor \cite{Karlof}, \cite{Haghighi}. } based on wormhole (tunneling) \cite{Hu}, \cite{Giannetsos}, or
compromised nodes \cite{Yoo}, \cite{Saghar}.

\subsubsection{Topology Discovery }

As described in Sec. \ref{sec:Introduction}, multicasting a topology
frame is impractical in WTDP. To solve this issue, we propose that,
in WTDP, the topology frame contains an ordered, rather than an unordered
as in wired TDP, list of MAC addresses in the current topology table
of the sender node. Specifically, the topology frame sent to the neighbor
on the right contains all the currently known MAC address of the BNs
on the left of the BN in the discovered physical order, and vice versa
for the topology frame sent to the neighbor on the left. The topology
frame also includes the CN IDs that are connected to, rather than
only the sender BN as in wired TDP, all BNs currently discovered.
Taking the wireless backbone network in Fig. \ref{fig. Wireless1}
as an example, if the associated logical topology is identical to
that shown in Fig. \ref{fig. wiredTrain}, and if BN 4 has identified
BN 3 as a neighbor and has discovered that BNs 5 and 6 are on its
right, the topology frame sent by BN 4 to BN 3 contains an ordered
list of MAC addresses of BNs 5 and 6. It also includes the IDs of
the CNs that are connected to the BNs 4, 5 and 6, namely, CN B.3 and
CN B.2 are connected to BN 4; CN B.3 is connected to BN 5; CN B.2
and CN B.1 are connected to BN 6. A topology frame is sent to an identified
or locked neighbor. Based on a received topology frame, the receiving
BN updates its local topology table only if the received topology
frame is from its locked neighbor. Note that after the physical topology
is learned, the logical topology can be learned in the same way as
in the wired TDP (see \cite{IEC 61375} for details). Therefore, we
focus on the physical topology discovery for WTDP next.

After a successful neighbor discovery has been resolved for all BNs,
it is necessary and sufficient to have a ``right-ward'' and a ``left-ward''
pass in order to complete topology discovery. For instance, in Fig.
\ref{fig. Wireless1}, assume that the protocol starts from BN 1,
which sends a topology frame to its neighbor on the right BN 2, which
in turn sends a topology frame to its neighbor BN 3, and so on until
BN 6. At the end of this right-ward pass, it is easy to see that each
BN in Fig. \ref{fig. Wireless1} learns the backbone topology on its
right. A similar left-ward pass completes the topology discovery at
each BN. It can also be seen that the mentioned frames are also necessary
in order to learn the train topology. 

We emphasize that the proposed WTDP differs from the standard wired
TDP in that the latter prescribes multicasting of topology frames
and the inclusion of an unordered list of discovered nodes and the
CN IDs that are connected to the sender BN only in the topology frames.
The operation of the topology discovery phase is described by the
functions enclosed in the ``topology discovery'' block of Fig. \ref{fig:flow}.

\subsubsection{Topology Convergency Check\label{sub:Topology-Convergency-Check}}

In order for the operator to make a decision about the completion
of the inauguration process, the BNs must report on the status of
their topology discovery phase. To this end, each BN runs a topology
convergence check as shown in the ``topology convergency check''
block of Fig. \ref{fig:flow}. Accordingly, when a topology frame
is received from a locked neighbor, if any change needs to be made
to the local topology table, the topology counter for the BN is initialized;
otherwise, the counter is increased by one. The topology discovery
completion for a BN is claimed if the topology counter reaches a pre-defined
threshold $M_{T}$. In other words, the topology discovery is considered
to be complete by a BN if no change is made to its topology frame
across $M_{T}$ successively received topology frames from the locked
neighbor. The completion of topology discovery is indicated by green
flags raised by the BNs.

\section{Performance Analysis\label{sec:Performance-Analysis-with}}

In order to get some insights into the performance of the proposed
WTDP, we consider the implementation of WTDP with a slotted ALOHA
MAC protocol. Note that the protocol does not depend on the adoption
of a specific MAC layer protocol and that slotted ALOHA is assumed
here to enable analysis. According to slotted ALOHA, time is slotted,
a transmitted frame takes one slot, and each BN transmits a frame
in a slot with probability $p$. Specifically, at each time slot,
a BN transmits a hello frame with probability $p_{H}$, and transmits
a topology frame with probability $p_{T}$. Hence, the transmission
probability $p$ is the sum of $p_{H}$ and $p_{T}$, i.e., $p=p_{H}+p_{T}$. 

Flat Rayleigh fading channels are assumed such that the instantaneous
channel gain between two nodes $k$ hops away can be written as $\mathrm{SNR_{0}}\left|h\right|^{2}/(1+(k-1)F)^{\eta}$,
where we define the average signal to noise ratio (SNR) for two nodes
one hop away as $\mathrm{SN}\mathrm{R}_{0}$, $|h|^{2}$ is exponentially
distributed with mean one, and $\eta$ denotes the path loss exponent.
Furthermore, the channels across different time slots are assumed
to be independent, while the channel is a constant within the period
of a frame transmission. We note that a more general channel model,
such as Rician or Nakagami fading, could also be accommodated in the
analysis but at the cost of a more cumbersome notation due to the
lack of some closed-form expressions that are available for Rayleigh
fading as discussed below. We present experiments with Rician fading
in Sec. \ref{sec:Numerical-Results-and}. 

In the following, we provide an analysis for the neighbor discovery
phase in terms of the probability of correct neighbor discovery and
of the average time required to complete neighbor discovery. There
two conflicting criteria will also be combined to yield the average
time needed to achieve successful neighbor discovery. The goal of
the analysis is to obtain insights into the selection of the critical
threshold parameter $M_{H}$. The performance of the overall WTDP
will be evaluated in the next section via numerical results.

\subsection{Neighbor Discovery for a Single BN }

In this subsection, we consider the neighbor discovery for a single
receiving BN on any given side. We compute the probability $Q_{C,ND}$
of correct neighbor discovery and the cumulative distribution function
(CDF) $F_{T_{ND}}\left(t\right)$ of the time $T_{ND}$ that it takes
to complete neighbor discovery. 

To elaborate, assume that the furthest BN from which hello frames
can be received is $K$ hops away. The signal-to-interference-and-noise
ratio (SINR) for the signal transmitted by a BN $k$ hops away is
given by 
\begin{equation}
\mathrm{SIN}\mathrm{R}_{k}=\frac{\left|h_{k}\right|^{2}\frac{\textrm{\ensuremath{\mathrm{SNR_{0}}}}}{(1+(k-1)F)^{\eta}}}{1+\sum_{k'\neq k}^{K}i_{k'}\left|h_{k'}\right|^{2}\frac{\textrm{\ensuremath{\mathrm{SNR}_{0}}}}{(1+(k'-1)F)^{\eta}}},\label{eq:SINR}
\end{equation}
where $i_{k'}=1$ if the BN $k'$ hops away is transmitting and $i_{k'}=0$
otherwise. Moreover, the instantaneous channel capacity for the link
between the two nodes, which are $k$ hops away from each other, is
given by \cite{cover} 
\begin{equation}
C_{k}=\log\left(1+\mathrm{SINR}_{k}\right).\label{eq:capacity}
\end{equation}
Whenever the transmission rate $\mathrm{\mathit{R\,}[bits/sec/Hz]}$
is not larger than the instantaneous capacity $C_{k}$, the packet
transmitted by the BN $k$ hops away is correctly received, and an
outage is declared otherwise \cite{Tse}. 

Define the vector $\mathbf{i}=[i_{1},\cdots,i_{K}]^{T}$ that defines
the set of currently transmitting BNs. At any time slot, the probability
of a successful frame reception from a node $k$ hops away conditioned
on $\mathbf{i}$ can be expressed as
\begin{align}
 & Q{}_{S}(k|\mathbf{i})=i_{k}\frac{p_{H}}{p}\Pr\left[C_{k}\geq R|\mathbf{i}\right].\label{eq:PCN_condition_1}
\end{align}
Substituting (\ref{eq:capacity}) into (\ref{eq:PCN_condition_1})
leads to
\begin{align}
Q{}_{S}(k|\mathbf{i}) & =i_{k}\frac{p_{H}}{p}\Pr\Bigg[\frac{|h_{k}|^{2}\mathrm{SNR_{0}}}{(1+(k-1)F)^{\eta}}\geq\nonumber \\
 & 2^{R}-1+\sum_{k'\neq k}^{K}i_{k'}(2^{R}-1)\mathrm{\frac{\mbox{\ensuremath{|h_{k'}|^{2}}}SNR_{0}}{(\mbox{\ensuremath{1+(k'-1)F}})^{\eta}}\Bigg]}.\label{eq:PCN_condition_2}
\end{align}
 Using the result in \cite{Boyd}, we get
\begin{align}
Q{}_{S}(k|\mathbf{i})= & i_{k}\frac{p_{H}}{p}\exp\left(-\frac{2^{R}-1}{\mathrm{SNR_{0}}}(1+(k-1)F)^{\eta}\right)\nonumber \\
\times & \prod_{k'\neq k}^{K}\left(\frac{1}{1+i_{k'}(2^{R}-1)\frac{(1+(k-1)F)^{\eta}}{(\mbox{\ensuremath{1+(k'-1)F}})^{\eta}}}\right).\label{eq:PCN_condition}
\end{align}
Averaging over all possible transmission states $\mathbf{i}$, we
can write the probability of a successful frame transmission from
a node $k$ hops away as
\begin{equation}
Q{}_{S}(k)=\sum_{\mathbf{i}\in\mathbf{\mathbf{\mathcal{I}}}}P_{\boldsymbol{\mathbf{i}}}\left(\mathbf{i}\right)Q{}_{S}(k|\mathbf{i}),\label{eq:PCN_inter}
\end{equation}
where $\mathbf{\mathcal{I}}$ denotes the set that contains all possible
$2^{K}$ transmission state vectors and $P_{\mathbf{\mathbf{i}}}\left(\mathbf{i}\right)$
is the probability mass function of vector $\mathbf{\mathbf{i}}$.

Due to the independence of the fading channels across the time slots,
the time $T_{k}$ that it takes to receive $M_{H}$ hello frames from
a BN $k$ hops away is distributed as $T_{k}\sim NB(M_{H},Q_{S}(k))$,
where we use the notation $NB(M,p)$ to denote a negative binomial
distribution\footnote{In a sequence of independent Bernoulli ($p$) trials, let the random
variable $N$ denote the trial at which the $M$th success occurs,
where $M$ is a fixed integer. Then $N$ has a negative binomial distribution
\cite{Morris} with parameter $(M,p)$, i.e.$N\sim NB(M,p)$. } with parameter $(M,p)$. Accordingly, the probability mass function
of $T_{k}$ is given by \cite{Morris} 
\begin{equation}
P_{T_{k}}\left(t\right)=\left(\begin{array}{c}
t-1\\
M_{H}-1
\end{array}\right)Q_{S}(k)^{M_{H}}(1-Q_{S}\left(k\right))^{t-M_{H}},
\end{equation}
$\textrm{for \ensuremath{t\geq M_{H}}}$; and the complementary cumulative
distribution function (CCDF) of $T_{k}$, which equals to the probability
that hello frames sent by a BN $k$ hops away are received successfully
$M_{H}$ times after the $t$th time slot, can be expressed as \cite{Morris}
\begin{equation}
\bar{F}_{T_{k}}(t)=1-I_{Q_{S}(k)}(M_{H},t-M_{H}+1)\textrm{, for \ensuremath{t\geq M_{H}}}.\label{eq:CCDF_oneNode}
\end{equation}
where $I_{x}(z,w)$ denotes the regularized incomplete beta function
with parameters ($x,z,w$).

So far, we have considered the distribution of the time needed to
receive $M_{H}$ hello frames from a given transmitting BN. We are
now interested in deriving the probability $Q_{C,ND}$ of correct
neighbor discovery. This calculation is complicated by the fact that
the receptions of frames from different BNs are correlated with each
other due to the mutual interference among BNs. To address this issue,
we make here the approximation that the decoding outcomes for the
packets sent by different BNs are independent. The validity of this
approximation will be evaluated in Sec. \ref{sec:Numerical-Results-and}
by numerical results. Recall that, if the first BN from which hello
frames are received successfully $M_{H}$ times is the BN one hop
away, neighbor discovery is correct. Hence, using the said independence
assumption, the probability of correct neighbor discovery for a single
receiving BN is 
\begin{equation}
Q_{C,ND}=\sum_{t\geq M_{H}}^{\infty}\left[P_{T_{1}}\left(t\right)\prod_{k\geq2}^{K}\bar{F}_{T_{k}}(t)\right].\label{eq:PCN_oneNode}
\end{equation}
Finally, regardless of whether it is correct or not, neighbor discovery
is considered to be complete when a BN decodes $M_{H}$ hello frames
successfully from at least one of other transmitting BNs. The CDF
of the time it takes to complete neighbor discovery for the BN $T_{ND}$
can be expressed, under the independence assumption, as 
\begin{align}
F_{T_{ND}}\left(t\right) & =1-\Pr\left[T_{ND}>t\right]\nonumber \\
 & =1-\Pr\left[\mathrm{\min}\left\{ T_{1},T_{2}...,T_{K}\right\} >t\right]\nonumber \\
 & =1-\prod_{k=1}^{K}\bar{F}_{T_{k}}(t).\label{eq:CDF_T_oneNode}
\end{align}

\subsection{Neighbor Discovery Across the Entire Network\label{par:ND noInter allBNs}}

In this subsection, we derive the performance metrics of neighbor
discovery across the entire network. Specifically, we derive the probability
$Q_{C,ND}^{*}$ of correct neighbor discovery, the average time $E[T_{ND}^{*}]$
required to complete neighbor discovery for all nodes and the average
time $E[T_{ND,\mathrm{\mathit{suc}}}^{*}]$ needed to achieve a successful
neighbor discovery. 

Because the neighbor discovery outcomes for different BNs are not
independent, the analytical derivation of statistical quantities associated
with neighbor discovery performance is challenging. For this reason,
we will make the approximation mentioned above that the neighbor discovery
outcomes for different BNs are independent. With this approximation,
probability of correct left and right neighbor discovery for all  BNs
in the network can be expressed as 
\begin{equation}
Q_{C,ND}^{*}=\left(Q_{C,ND}\right)^{2D},\label{eq:PCN}
\end{equation}
where $D$ denotes the total number of receiving BNs and the factor
$2$ stems from the fact that different frequencies are used for transmission
and reception and hence, the left neighbor discovery is independent
from the right neighbor discovery. Similarly, the CDF of the time
it takes to achieve a successful neighbor discovery on both left and
right sides is given by 
\begin{align}
F_{T_{ND,L}^{*}}\left(t\right) & =\left(F_{T_{ND,L}}\left(t\right)\right)^{2D}.\label{eq:CDF_T_L}
\end{align}
The average time needed to achieve a successful neighbor discovery
$E[T_{ND}^{*}]$ is then given by 
\begin{equation}
E\left[T_{ND}^{*}\right]=\sum_{t=0}^{\infty}\left[1-F_{T_{ND}^{*}}(t)\right].\label{eq:AvgTime}
\end{equation}

Next we combine the two statistical quantities $Q_{C,ND}^{*}$ and
$E\left[T_{ND}^{*}\right]$ to yield the average time it takes to
achieve a successful neighbor discovery $E[T_{ND,\mathrm{\mathit{suc}}}^{*}]$.
Using Wald's equality \cite{Janssen}\footnote{If $\{X_{n};\,n\geq1\}$ is a sequence of independent identically
distributed random variables with mean $\overline{X}$ and if the
mean $E[J]$ of the stopping time $J$ satisfies $E[J]<\infty$, then
the sum $S_{J}=X_{1}+X_{2}+\cdots+X_{J}$ at the stopping time $J$
satisfies Wald's equality $E\left[S_{J}\right]=\overline{X}E\left[J\right]$. }, this can be evaluated as the ratio 
\begin{equation}
E[T_{ND,suc}^{*}]=\frac{E\left[T_{ND}^{*}\right]}{Q_{C,ND}^{*}}.\label{eq:AvgStime}
\end{equation}

\section{Case Study: Trains on Parallel Tracks\label{sec:ND-parellel_tracks}}

\begin{figure}[htbp]
\begin{centering}
\textsf{\includegraphics[width=7.2cm,height=2cm]{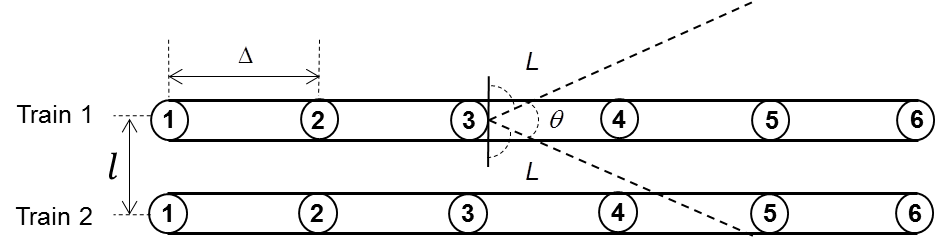}}
\par\end{centering}

\caption{\label{fig:parellel tracks} An example of two trains on parallel
tracks (top view). Also shown is the antenna gain pattern for BN 3
on train 1.}
\end{figure}

In this section, we describe a scenario of practical interest in which
two trains located on parallel tracks perform separate inauguration
processes. This set-up will be further elaborated on in Sec. \ref{sec:Numerical-Results-and}
via numerical results. As shown in Fig. \ref{fig:parellel tracks},
we assume the same number of BNs for each train, and we denote the
distance between a BN and its neighbor on the same train as $\Delta$,
while $l$ denotes the distance between two trains. We also assume
that BNs are aligned as in Fig. \ref{fig:parellel tracks}. Let the
directional antenna of each BN have a mainbeam of width $\theta$,
while sidelobes have a $L$ dB loss compared to the mainlobe. For
instance, in Fig. \ref{fig:parellel tracks}, BN 3 and BN 4 on train
2 are in the side lobe region of the right-pointing antenna of BN
3 on train 1, and hence are received by BN 3 on train 1 with a loss
of $L$ dB. Instead, no loss occurs for the reception by BN 3 on train
1 of the signals sent by BN 5 and BN 6 on train 2 or BNs 4-6 on train
1. 

\section{Numerical Results and Discussions\label{sec:Numerical-Results-and}}

In this section, we evaluate the performance of the proposed WTDP
as applied to a wireless network that runs the ALOHA MAC protocol.
Unless stated otherwise, we assume the following conditions: \textit{i})
a flat Rayleigh fading channels as described in Sec. \ref{sec:Performance-Analysis-with};
\textit{ii}) a path loss exponent $\eta=3.5$; \textit{iii}) an average
SNR of $15\,\mathrm{dB}$ for two nodes one hop away, i.e., $\mathrm{SN}\mathrm{R}_{0}=15\,\mathrm{dB}$;
\textit{iv}) a total of six BNs in the network; \textit{v}) at any
time slot, a hello frame is transmitted with probability, $p_{H}=0.15$,
and a topology frame is transmitted with probability, $p_{T}=0.15$;
\textit{vi}) a data rate $R$ $1.5\,\mathrm{bits/sec/Hz}$ for the
hello frames, and \textit{vii}) full frequency reuse is adopted, i.e.,
$F=1$. We recall that a more conservative frequency reuse would alleviate
interference and therefore improve the performance. 
\begin{figure}[h]
\begin{centering}
\textsf{\includegraphics[width=8.5cm,height=5.95cm]{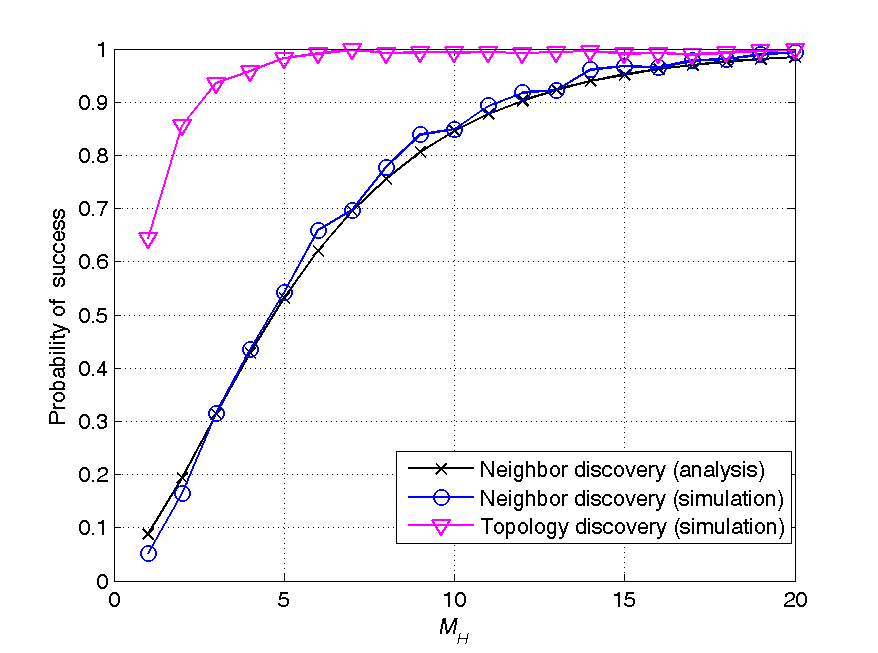}}
\par\end{centering}

\caption{\label{fig: PvsM_H}The probability of success of neighbor discovery
and of overall inauguration process versus the threshold $M_{H}$
used for neighbor discovery ($p_{H}=0.15$, $p_{T}=0.15$ and $\mathrm{SN}\mathrm{R}_{0}=15\,\mathrm{dB}$).}
\end{figure}
\begin{figure}[h]
\begin{centering}
\textsf{\includegraphics[width=8.5cm,height=5.95cm]{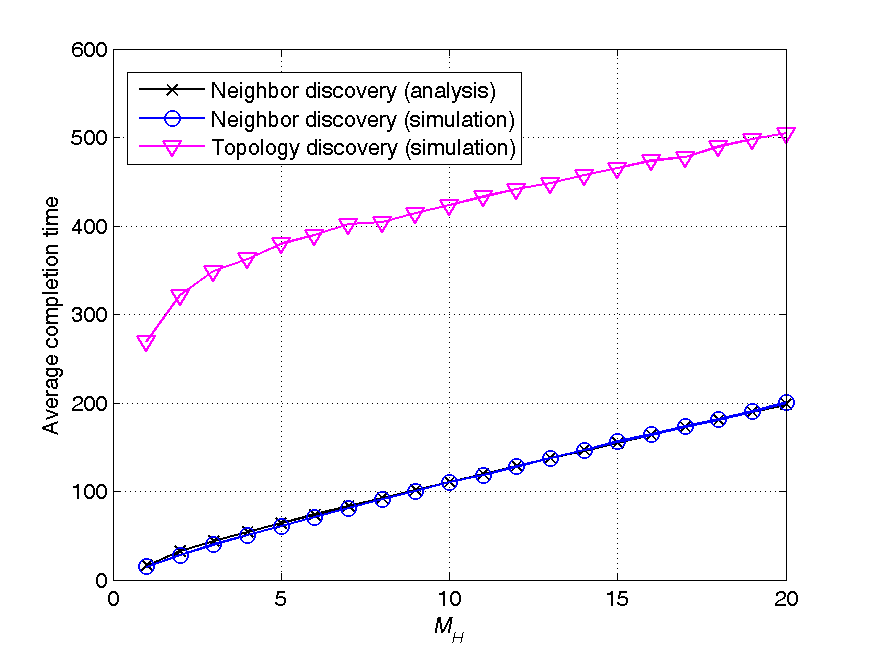}}
\par\end{centering}

\caption{\label{fig: TvsM_H}The average time required to complete neighbor
discovery and to complete the entire inauguration process versus $M_{H}$
($p_{H}=0.15$, $p_{T}=0.15$ and $\mathrm{SN}\mathrm{R}_{0}=15\,\mathrm{dB}$).}
\end{figure}
\begin{figure}[h]
\begin{centering}
\textsf{\includegraphics[width=8.5cm,height=5.95cm]{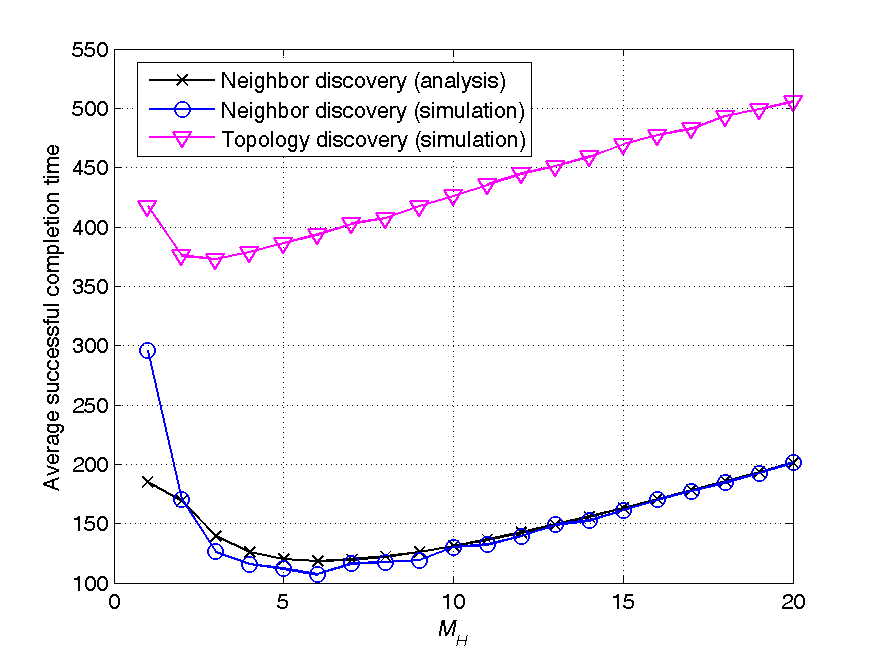}}
\par\end{centering}

\caption{\label{fig: T_SvsM_H}The average time needed to achieve a successful
neighbor discovery and overall inauguration versus $M_{H}$ ($p_{H}=0.15$,
$p_{T}=0.15$ and $\mathrm{SN}\mathrm{R}_{0}=15\,\mathrm{dB}$).}
\end{figure}

\subsection{Effects of the Threshold $M_{H}$}

We first investigate the effect of the threshold parameter $M_{H}$
on the neighbor discovery performance. This discussion is also meant
to corroborate the validity of the analysis in Sec. \ref{sec:Performance-Analysis-with}.
In Figs. \ref{fig: PvsM_H}-\ref{fig: T_SvsM_H}, the probability
of correct neighbor discovery, the average time required to complete
neighbor discovery and the average time needed to achieve a successful
neighbor discovery are shown as a function of $M_{H}$, respectively.
We plot both the analytical results (\ref{eq:PCN}), (\ref{eq:AvgTime})
and (\ref{eq:AvgStime}) and the performance obtained via Monte Carlo
simulations. It can be seen from Figs. \ref{fig: PvsM_H}-\ref{fig: T_SvsM_H},
that the analysis predicts the performance of neighbor discovery well
in terms of the three criteria. As expected, the success rate and
time needed to complete neighbor discovery increase as threshold $M_{H}$
increases. This leads to a trade-off in the selection of $M_{H}$:
a larger $M_{H}$ improves the probability of successful neighbor
discovery but, at the same time, it increases the time needed for
neighbor discovery. This trade-off is illustrated in Fig. \ref{fig: T_SvsM_H},
which demonstrates that there exists a value of $M_{H}$ that minimizes
the time needed to achieve successful neighbor discovery. We observe
that the analysis allows to correctly predict the optimal value of
$M_{H}$. 

In Figs. \ref{fig: PvsM_H}-\ref{fig: T_SvsM_H}, the performance
for the overall proposed inauguration process including all the phases
described in Sec. \ref{sec:Proposed-Inauguration-Scheme} is also
presented. To this end, we set $M_{NDF}=20$ and $M_{T}=30$ and evaluate
the performance via Monte Carlo simulations. The dramatic success
rate improvement for the inauguration over neighbor discovery is to
be ascribed to the PCC. This improvement can be also seen to decrease
the optimal value of $M_{H}$. It can also be observed that there
is a difference of about 300 time slots between the time required
to complete neighbor discovery and the time required to complete the
whole inauguration process. This is due to the fact that besides neighbor
discovery, the inauguration process needs to complete also topology
discovery. 

\begin{figure}[h]
\begin{centering}
\textsf{\includegraphics[width=8.5cm,height=5.95cm]{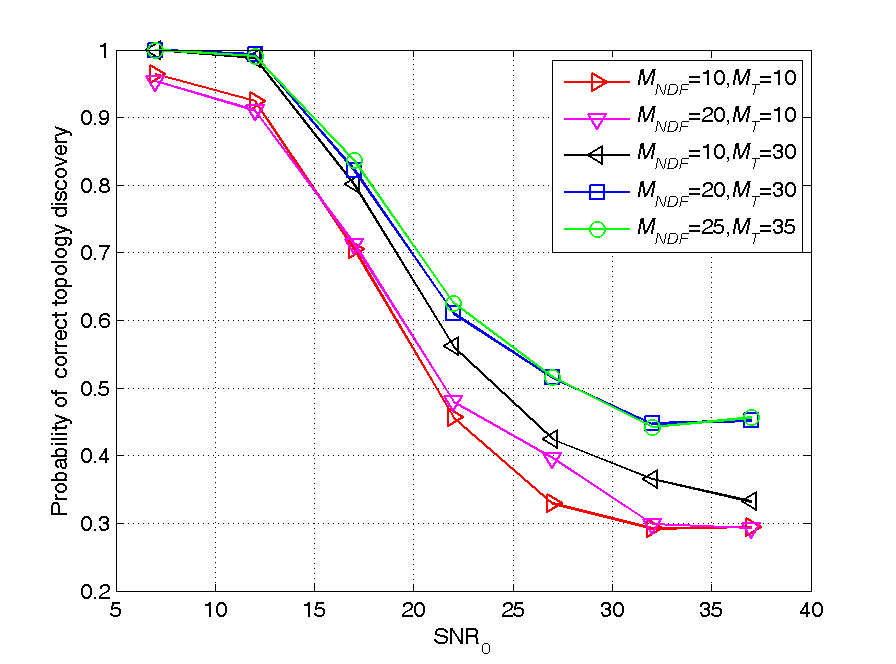}}
\par\end{centering}

\caption{\label{fig: PvsSNR}The probability of successful topology discovery
versus $\mathrm{SN}\mathrm{R}_{0}$ for different values of the thresholds
$M_{NDF}$ and $M_{T}$ ($p_{H}=0.15$, $p_{T}=0.15$ and $M_{H}=3$).}
\end{figure}
\begin{figure}[h]
\begin{centering}
\textsf{\includegraphics[width=8.5cm,height=5.95cm]{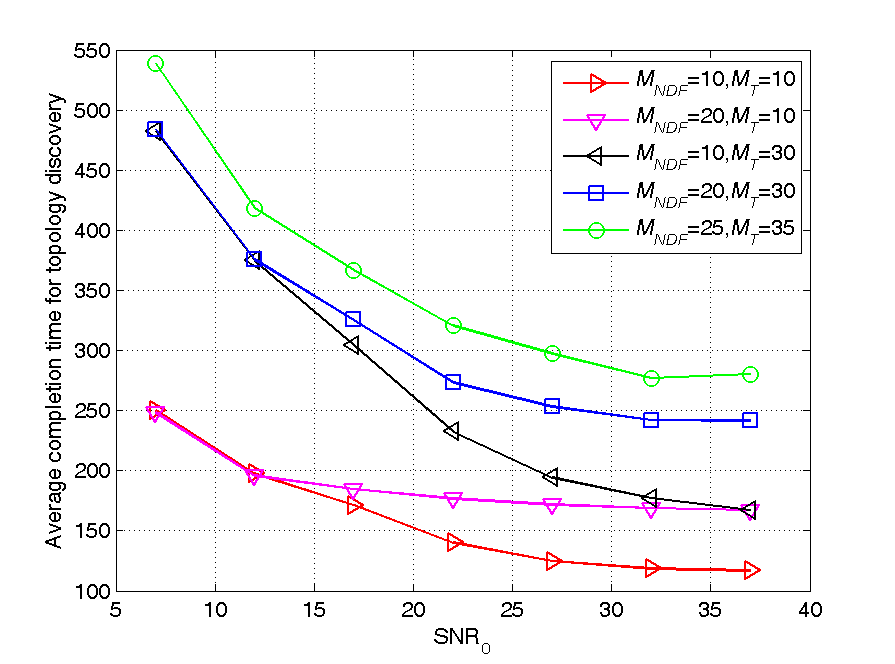}}
\par\end{centering}

\caption{\label{fig:TvsSNR}The average time required to complete topology
discovery vs. $\mathrm{SNR}_{0}$ for different values of the thresholds
$M_{NDF}$ and $M_{T}$ ($p_{H}=0.15$, $p_{T}=0.15$ and $M_{H}=3$).}
\end{figure}

\subsection{Effects of Thresholds $M_{NDF}$ and $M_{T}$ }

We now explore the effects of two thresholds $M_{NDF}$ and $M_{T}$
on the performance of WTDP. The value of threshold $M_{H}$ is set
to 3 based on the discussion above. In Fig. \ref{fig: PvsSNR}, we
plot the probabilities of correct topology discovery and in Fig. \ref{fig:TvsSNR}
we show the average time required to complete topology discovery versus
the average one-hop SNR, parameterized by different values of $M_{NDF}$
and $M_{T}$. It can be seen that larger thresholds $M_{NDF}$ and
$M_{T}$ result in an improved probability of correct topology discovery.
This is because it is more unlikely that an incorrect identification
of neighbor discovery failure occurs with a larger $M_{NDF}$ (see
Sec. \ref{sub:NDFC}) while a larger $M_{T}$ tends to improve the
efficiency of the topology convergency check (see Sec. \ref{sub:Topology-Convergency-Check}).
On the flip side, Fig. \ref{fig:TvsSNR} shows that larger values
of $M_{NDF}$ and $M_{T}$ always lead to longer average time needed
to complete the inauguration.

\subsection{Neighbor Discovery Over Rician Fading Channels\label{sub:Neighbor-Discovery-Over}}

\begin{figure}[h]
\begin{centering}
\textsf{\includegraphics[width=8.5cm,height=5.95cm]{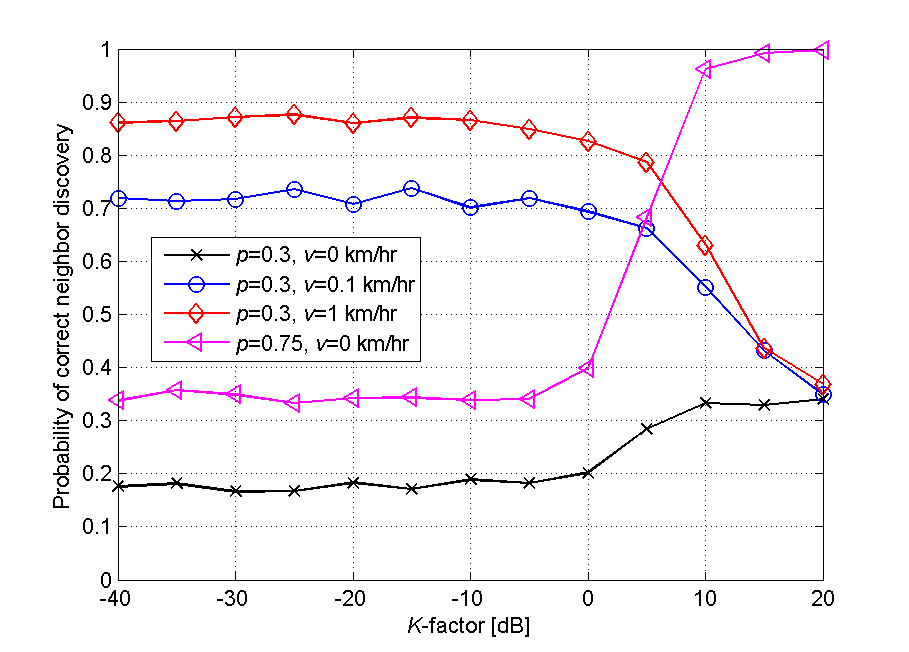}}
\par\end{centering}

\caption{\label{fig:Rician}The probability of successful neighbor discovery
versus the $K$-factor for Rician fading channels with different values
of train speeds and the transmission probabilities ($M_{H}=10$ and
$\mathrm{SNR}_{0}=15\,\mathrm{dB}$). }
\end{figure}

We now consider neighbor discovery over flat-fading Rician channels.
We recall that the defining parameter of Rician fading is the $K$-factor,
which is defined as the power ratio of the line-of-sight component
and diffuse components \cite{Rappaport}. In Fig. \ref{fig:Rician},
we present the probability of correct neighbor discovery versus the
Rician $K$-factor with different values of the transmission probability
$p$ and of the train speed $v$. We adopt the standard Jakes model
\cite{Rappaport} to account for channel correlation as a function
of the train velocity v. We set the threshold for neighbor discovery
to $M_{H}=10$ and equal probability for transmission of a hello and
a topology frames. It can be seen from Fig. \ref{fig:Rician} that
in the low-$K$ regime, the success rate is low over static channels,
i.e. $v=0\,\mathrm{km/hour}$, but a minor increase in the train speed,
i.e., with $v=1\,\mathrm{km/hour}$, significantly improves the success
rate. This is because with static channels, time diversity is lost,
but due to the long duration of a time slot ($T=100\,\mathrm{ms}$),
a speed as low as $1\,\mathrm{km/hour}$ results in uncorrelated channel
gains across different time slots. This can be verified by the fact
that the success rate with low $K$-factor at the speed of $1\,\mathrm{km/hour}$
(see Fig. \ref{fig:Rician}) converges to the success rate of 86\%,
which is also the success rate for neighbor discovery with the threshold
$M_{H}=10$ shown in Fig. \ref{fig: PvsM_H}. Instead, in high-$K$
regime, a larger transmission probability $p$ results in higher probability
correct neighbor discovery  in the static case. This is explained
by the fact that in this regime, the channel gain tends to be dominant
by the line-of-sight component, yielding successful frame transmissions
from both neighboring BNs and non-neighboring BNs in absence of collision.
A larger transmission probability results in more collisions, which
in turn reduce the chance of successful frame decoding, more severely
for frames sent by non-neighboring BNs than for the ones sent by neighboring
BNs, since the latter BNs are received with sufficient power not to
incur outage. 

\subsection{Neighbor Discovery of Two Parallel Trains}

In this subsection, we evaluate the neighbor discovery performance
with two trains located on parallel tracks as described in Sec. \ref{sec:ND-parellel_tracks}.
The BNs of both trains transmit by using the slotted ALOHA protocol
as per Sec. \ref{sec:Performance-Analysis-with}. Note that while
this assumes synchronization between the trains, we expect the effect
of inter-train interference to be qualitatively the same even under
asynchronous MACs. We evaluate the neighbor discovery performance
for train 1 with train 2 serving as interference. Each train is equipped
with six BNs. The beam width $\theta$ is selected as $\theta=\pi/3$.
Rayleigh fading is assumed. 

\begin{figure}[h]
\begin{centering}
\textsf{\includegraphics[width=8.5cm,height=5.95cm]{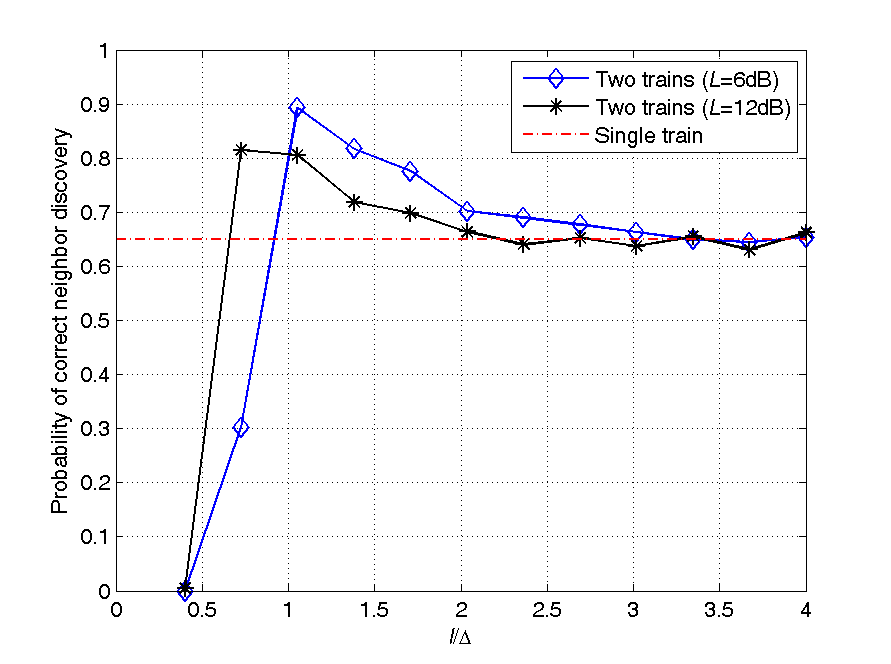}}
\par\end{centering}

\caption{\label{fig:PvsC}The probability of successful neighbor discovery
versus the ratio $l/\Delta$ (see Fig. \ref{fig:parellel tracks})
for the two-train set-up of Fig. \ref{fig:parellel tracks} with different
values of the side lobe attenuation $L$ ($p_{H}=0.15$, $p_{T}=0.15$,
$\mathrm{SNR}_{0}=15\,\mathrm{dB}$, $M_{H}=6$ and $\theta=\pi/3$).}
\end{figure}

\begin{figure}[h]
\begin{centering}
\textsf{\includegraphics[width=8.5cm,height=5.95cm]{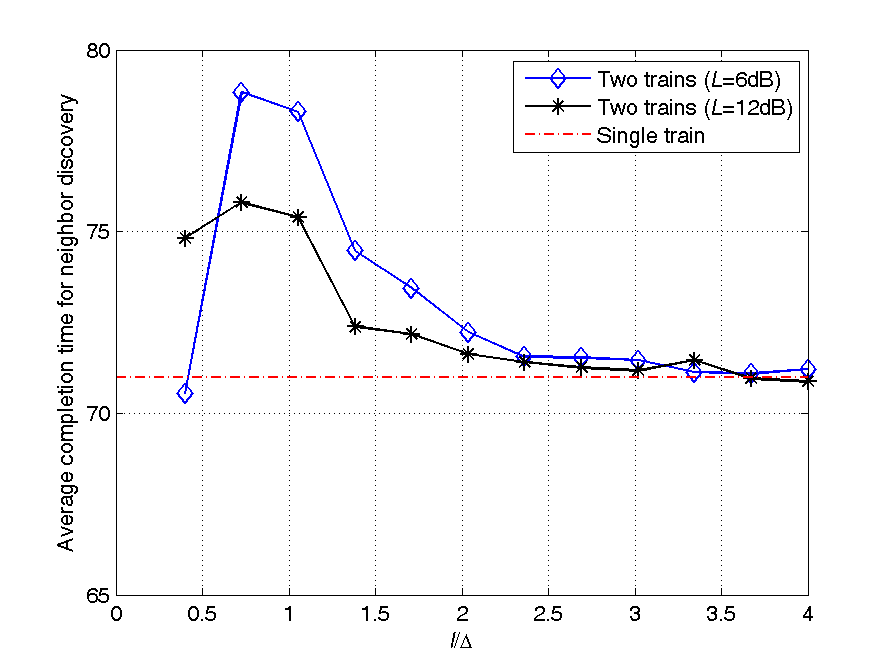}}
\par\end{centering}

\caption{\label{fig:TvsC}The average time required to complete neighbor discovery
versus the ratio $l/\Delta$ (see Fig. \ref{fig:parellel tracks})
for the two-train set-up of Fig. \ref{fig:parellel tracks} with different
values of the side lobe attenuation $L$ ($p_{H}=0.15$, $p_{T}=0.15$,
$\mathrm{SNR}_{0}=15\,\mathrm{dB}$, $M_{H}=6$ and $\theta=\pi/3$). }
\end{figure}

In Figs. \ref{fig:PvsC} and \ref{fig:TvsC}, we show the probability
of correct neighbor discovery and the average time needed to complete
neighbor discovery versus the ratio $l/\Delta$, parameterized by
sidelobe attenuation $L=6\,\mathrm{d\mathrm{B}}$ and $L=12\,\mathrm{dB}$.
Also shown for reference is the performance for the case in which
only train 1 is present, i.e., no inter-train interference exists.
It can be seen from Fig. \ref{fig:PvsC} that the accuracy of neighbor
discovery is poor for small $l/\Delta$, and that, as the ratio $l/\Delta$
increases, the probability of successful neighbor discovery first
increases and then decreases, reaching the interference-free performance
for $l/\Delta$ large enough. This can be explained as follows. With
small $l/\Delta$, the BNs on train 2 tend to be selected by the neighbor
discovery process run at BNs on train 1, causing the failure of neighbor
discovery. This effect becomes less pronounced as the ratio $l/\Delta$
increases and hence the performance is enhanced. Interestingly, for
values of $l/\Delta$ close to one, the interference may be even beneficial
to neighbor discovery. The reason for this is similar to the one for
the scenario in which concurrent transmission happens with a single
train (see Sec. \ref{sub:Neighbor-Discovery-Over}). It is also seen
that a larger sidelobe attenuation causes this effect to be observed
for lower values of $l/\Delta$. As $l/\Delta$ increases further,
the performance converges to that of a single train with no interference. 

In contrast to the probability of correct neighbor discovery, the
average time required to complete neighbor discovery is shown in Fig.
\ref{fig:TvsC} to be first degraded as $l/\Delta$ increases before
finally converging to the interference-free performance. This is because
when $l/\Delta$ is close to one, frames from both trains tend to
be received with similar powers leading to numerous outage events.
Instead, if $l/\Delta$ is smaller, the BNs on train 1 will more likely
choose BNs on train 2 as neighbors, while for larger $l/\Delta$,
neighbors tends to be successful.

\section{Concluding Remarks}

Topology discovery in wireless linear networks is a key enabling protocol
for application such as wireless train backbone communication. In
this work, we have presented a wireless topology discovery protocol
(WTDP) that requires minor modification to the current wired topology
discovery protocol standard and is able to cope with wireless impairments,
such as broadcasting, interference and fading. The proposed WTDP is
analyzed under a slotted ALOHA MAC protocol and shows, with the aid
of extensive numerical examples, to provide flexible and robust performance
under realistic condition including the case of inter-train interference.
Interesting future work includes the investigation of the impact of
more sophisticated physical layer technologies, such as MIMO, on topology
discovery, a more thorough analysis of the effect of fast fading channels,
as well as the study of privacy and security issues associated with
wireless topology discovery.

\section*{Acknowledgment}

We would like to thank Shahrouz Khalili for his help in the early
shape of this work with the wired standard \cite{IEC 61375}.


\begin{thebibliography}{10}
\bibitem{Kesting}A. Kesting, M. Treiber and D. Helbing, ``Connectivity
statistics of store-and-forward intervehicle communication,''\textit{
IEEE Trans. Intell. Transp. Syst.}, vol. 11, no. 1, pp. 172\textendash 181,
Mar. 2010.

\bibitem{Chakravarth}A. Chakravarthy, K. Song and E. Feron, \textquotedblleft Preventing
automotive pileup crashes in mixed-communication environments,\textquotedblright{}
\textit{IEEE Trans. Intell. Transp. Syst.}, vol. 10, no. 2, pp. 211\textendash 225,
Jun. 2009.

\bibitem{Arem}B. van Arem, C. van Driel and R. Visser, \textquotedblleft The
impact of cooperative adaptive cruise control on traffic-flow characteristics,\textquotedblright{}
\textit{IEEE Trans. Intell. Transp. Syst.}, vol. 7, no. 4, pp. 429\textendash 436,
Dec. 2006.

\bibitem{Hannes}H. Hartenstein, B. Bochow, A. Ebner, M. Lott, M.
Radimirsch and D. Vollmer, ``Position-aware ad hoc wireless networks
for inter-vehicle communications: the Fleetnet project,'' in \textit{Proc.
2nd ACM international symposium on Mobile ad hoc networking \& computing},
pp. 259-262, Long Beach, CA, Oct. 2001.

\bibitem{Biswas}S. Biswas, R. Tatchikou and F. Dion, ``Vehicle-to-vehicle
wireless communication protocols for enhancing highway traffic safety,''
\textit{IEEE Communications Magazine}, vol. 44, no. 1, pp. 74-82,
Jan. 2006. 

\bibitem{CHen}Z. D. Chen, H. T. Kung and D. Vlah, ``Ad hoc relay
wireless networks over moving vehicles on highways,'' in \textit{Proc.
2nd ACM international symposium on Mobile ad hoc networking \& computing},
pp. 247-250, Long Beach, CA, Oct. 2001.

\bibitem{Ning}B. Ning, T. Tang, Z. Y. Gao, F. Yan, F. Y. Wang and
D. Zeng, \textquotedblleft Intelligent railway system in China,\textquotedblright{}
\textit{IEEE Trans. Intell. Transp. Syst.}, vol. 21, no. 5, pp. 80\textendash 83,
Sep. 2006.

\bibitem{IEC 61375}IEC 61375-2-5, ``Electronic railway equipment-Train
communication network-Part 2-5. Ethernet Train Backbone,'' Jan. 2012.

\bibitem{Zeng}W. Zeng, S. Vasudevan, X. Chen, B. Wang, A. Russell
and W. Wei, ``Neighbor discovery in wireless networks with multipacket
reception,'' in \textit{Proc. ACM International Symposium on Mobile
Ad Hoc Networking and Computing}, pp. 3, Paris, France, May 2011.

\bibitem{YFayyaz}Y. Fayyaz, M. Nasim M.Y. Javed, ``Maximal weight
topology discovery in ad hoc wireless sensor networks,'' in \textit{proc.
IEEE International Conference on Computer and Information Technology},
pp. 715-722, Bradford, UK, June 2010.

\bibitem{MDSarr}M. D. Sarr, F. Delobel, M. Misson and I. Niang, ``Automatic
discovery of topologies and addressing for linear wireless sensors
networks,'' in \textit{proc. IEEE Wireless Days}, pp. 1-7, Dublin,
Ireland, Nov. 2012.

\bibitem{TKontos}T. Kontos, G.S. Alyfantis, Y. Angelopoulos and S.
Hadjiefthymiades, ``A topology inference algorithm for wireless sensor
networks,'' in\textit{ proc. IEEE Symposium on Computers and Communications},
pp. 479-484, Cappadocia, Turkey, July 2012.

\bibitem{Bli}B. Li, W. Feng, L. Zhang and C. Spanos, ``DEPEND: density
adaptive power efficient neighbor discovery for wearable body sensors,''
in \textit{proc. IEEE International Conference on Automation Science
and Engineering}, pp. 581-586, Madison, WI, Aug. 2013.

\bibitem{MNasim}M. Nasim, Y. Fayyaz, M.Y. Javed, ``Bounded degree
energy aware topology discovery in ad hoc wireless sensor networks,''
in \textit{proc. International Conference on Intelligent Sensors,
Sensor Networks and Information Processing}, pp.13-18, Melbourne,
VIC. Dec. 2009.

\bibitem{ABarnawi}A. Barnawi, R. Hafez, ``A time \& energy efficient
topology discovery and scheduling protocol for wireless sensor networks,''
in \textit{proc. International Conference on Computational Science
and Engineering}, vol. 2, no. 1, pp. 570-578, Vancouver, BC, Aug.
2009.

\bibitem{Hermann}S. Hermann, ``Investigation of IEEE 802.11k-based
access point coverage area and neighbor discovery,'' in \textit{Proc.
IEEE Conf. Local Comp. Networks}, pp. 949-954, Oct. 2007. 

\bibitem{Mcglynn}M. J. McGlynn and S. A. Borbash, \textquotedblleft Birthday
protocols for low energy deployment and flexible neighbor discovery
in ad hoc wireless networks,\textquotedblright{} in \textit{Proc.
ACM MobiHoc}, pp. 137-145, Long Beach, California, Oct. 2001.

\bibitem{HXie}H. Xie, F. Zeng, P.Wang and Y. Yang, ``Research and
implementation of fast topology discovery algorithm for Zigbee wireless
sensor network,'' in \textit{Proc.} \textit{IEEE International Conference
on Electronic Measurement \& Instruments}, vol. 2, no. 1, pp.914-918,
Harbin, China, Aug. 2013.

\bibitem{LiuX}X. Liu, B. Li, S. Huang and M. Chen, ``A ZigBee wireless
sensor network topology discovery algorithm,'' \textit{Computer Engineering},
vol. 38, no. 4, Feb. 2012. 

\bibitem{IJawhar}I. Jawhar, X. Li, J. Wu and N. Mohamed, ``Backbone
discovery in thick wireless linear sensor networks,'' in \textit{proc.
IEEE International Conference on Mobile Ad Hoc and Sensor Systems},
pp. 606-611, Philadelphia, PA, Oct. 2014.

\bibitem{Jeon}J. Jeon, A. Ephremides, \textquotedblleft Neighbor
discovery in a wireless sensor network: multipacket reception capability
and physical-layer signal processing,\textquotedblright \textit{Journal
of Communications and Networks}, vol. 14, no. 5, Oct. 2012.

\bibitem{Borbash}S. A. Borbash, A. Ephremides and M. J. McGlynn,
\textquotedblleft An asynchronous neighbor discovery algorithm for
wireless sensor networks,\textquotedblright{} \textit{Ad Hoc Netw}.,
vol. 5, no. 8, pp. 998\textendash 1016, 2007.

\bibitem{Vasudevan}S. Vasudevan, D. Towsley, D. Goeckel and R. Khalili,
\textquotedblleft Neighbor discovery in wireless networks and the
coupon collector\textquoteright s problem,\textquotedblright{} in
\textit{Proc. ACM MobiCom}, pp. 181-192, Beijing, China, Sept. 2009.

\bibitem{Kurose}S. Vasudevan, J. Kurose, and D. Towsley, \textquotedblleft On
neighbor discovery in wireless networks with directional antennas,\textquotedblright{}
in \textit{Proc. IEEE INFOCOM}, vol. 4, pp. 2502-2512, Miami, Florida,
Mar. 2005.

\bibitem{MZhang}M. Zhang, M. Chan and A.L. Ananda, ``Location-aided
topology discovery for wireless sensor networks,'' in \textit{Proc.
IEEE International Conference on Communications}, pp. 2718-2722, Beijing,
China, May 2008.

\bibitem{B-ning-1}H. Wang, F. Richard Yu, L. Zhu, T. Tang, and B.
Ning, \textquotedblleft A Cognitive Control Approach to Communication-based
Train Control (CBTC) Systems,\textquotedblright{} \textit{IEEE Trans.
Intelligent Transportation Systems}, vol. 16, no. 4, pp. 1676-1689,
Aug. 2015.

\bibitem{B-ning-2}L. Zhu, F. Richard Yu, B. Ning, and T. Tang, \textquotedblleft Communication-Based
Train Control (CBTC) Systems with Cooperative Relaying: Design and
Performance Analysis,\textquotedblright{} \textit{IEEE Trans. Veh.
Tech.}, vol. 63, no. 5, pp. 2162-2172, Jun. 2014.

\bibitem{Karlof}C. Karlof and D. Wagner, \textquotedblleft Secure
routing in sensor networks: attacks and countermeasures,\textquotedblright{}\textit{
Ad hoc Networks}, vol. 1, no. 2, pp. 293\textendash 315, 2003.

\bibitem{Haghighi}M. S. Haghighi, K. Mohamedpour, V. Varadharajan
and B. G. Quinn, ``Stochastic modeling of hello flooding in slotted
CSMA/CA wireless sensor networks,'' \textit{IEEE Transactions on
Information Forensics and Security}, vol. 6, no. 4, Dec. 2011.

\bibitem{Hu}Y. C. Hu, A. Perrig, and D. B. Johnson, \textquotedblleft Wormhole
attacks in wireless networks,\textquotedblright{}\textit{ IEEE J.
Sel. Areas Commun.}, vol. 24, no. 2, pp. 370\textendash 380, Feb.
2006.

\bibitem{Giannetsos}T. and T. Dimitriou, ``LDAC: A localized and
decentralized algorithm for efficiently countering wormholes in mobile
wireless networks,'' \textit{Journal of Computer and System Sciences},
Vol. 80, no. 3, pp. 618-643, May 2014.

\bibitem{Yoo}S. Yoo, S. Kang, and J. Kim, \textquotedblleft SERA:
a secure energy reliability aware data gathering for sensor networks,\textquotedblright{}
\textit{Multimedia Tools Applicat.}, vo.1, no. 2, pp. 1\textendash 30,
Jan. 2011.

\bibitem{Saghar}K. Saghar, W. Henderson, and D. Kendel, ``Formal
modelling and analysis of routing protocol security in wireless sensor
networks,'' in \textit{proc. Annual Postgraduate Symposium on the
Convergence of Telecommunications}, pp. 179\textendash 184, Liverpool,
June 2009.

\bibitem{cover}T. M. Cover and J. A. Thomas, \textit{Elements of
Information Theory}, John Wiley \& Sons, 2006.

\bibitem{Tse}D. Tse and P. Viswanath, \textit{Fundamentals of Wireless
Communication}, Cambridge University Press, 2005.

\bibitem{Boyd}S. Kandukuri and S. Boyd, ``Optimal power control
in interference-limited fading wireless channels with outage-probability
specifications,'' \textit{IEEE Trans. Wireless Commun.}, vol. 1,
no. 1, pp. 46-55, Jan. 2002.

\bibitem{Morris}D. H. Morris and S. J. Mark, \textit{Probability
and Statistics}, Pearson, 2011.

\bibitem{Janssen}J. Janssen and R. Manca, \textit{Applied Semi-Markov
Processes}, Springer, 2006.

\bibitem{Rappaport}T. S. Rappaport, \textit{Wireless Communications:
Principles and Practice}, Prentice Hall, 2002.\end{thebibliography}
\end{document}